# Core-periphery Segregation in Evolving Prisoner's Dilemma Networks

Yunkyu Sohn[1,2], Jung-Kyoo Choi[3], T.K. Ahn[4]*

[1] Department of Political Science, University of California, San Diego, CA 92093, USA.

[2] Mechanical Engineering Research Institute and Division of Web Science and Technology, KAIST, Daehak-ro, Dajeon 305-701, Korea.

[3] School of Economics and Trade, Kyungpook National University, 1370 Sankyuk-dong, Buk-gu, Daegu 702-701, Korea.

[4] Department of Political Science and International Relations, Seoul National University, Gwanakro 1, Seoul 151-746, Korea.

*To whom correspondence should be addressed. E-mail: tkahn@snu.ac.kr

**Dense cooperative networks are an essential element of social capital for a prosperous society. These networks enable individuals to overcome collective action dilemmas by enhancing trust (*1-3*). In many biological and social settings, network structures evolve endogenously as agents exit relationships and build new ones (*4-6*). However, the process by which evolutionary dynamics lead to self-organization of dense cooperative networks has not been explored. Our large group prisoner's dilemma experiments with exit and partner choice options show that core-periphery segregation of cooperators and defectors drives the emergence of cooperation. Cooperators' Quit-for-Tat (*7-10*) and defectors' Roving strategy (*11*) lead to a highly asymmetric core and periphery structure. Densely connected to each other, cooperators successfully isolate**

**defectors and earn larger payoffs than defectors. Our analysis of the topological characteristics of evolving networks illuminates how social capital is generated.**

Biological and social agents achieve cooperation through diverse mechanisms (*12-15*). Preferential association is one of these mechanisms in settings where agents are not forced to interact with fixed partners (*16-19*). Preferential association characterizes a wide range of domains from coauthorship among researchers and joint ventures among business firms to free-trade agreements among nation states in which agents can build and break interaction links with other agents (*20*). Several theoretical studies show that partner selection promotes cooperation by allowing cooperators to interact with other cooperators (*10, 21*). Available evidence from experiments using human subjects (*17, 22*) also supports the optimistic scenario.

However, the coevolutionary process between agents' strategic behavior and network structure is not yet understood. For example, due to the small number of subjects and the fixed number of links per subject, previous experimental studies were not able to trace the macroscopic coevolution of strategies and networks in the real world that consist of agents with extremely heterogeneous topological traits (*20*). In addition, the characteristics of individual level networking and gaming strategies that drive the coevolutionary process has not been investigated experimentally. To fill this gap in knowledge we ran experiments in which agents can break existing relationships and build new ones, enter multiple relationships and play cooperatively with some partners but opportunistically against others. We used groups of 35 to 40 subjects per session and allow subjects to freely re-establish relationships. The experiments were conducted in two different link cost settings to examine the robustness of the evolving network patterns.

In our experiments, multiple subjects played 2-person Prisoner's Dilemma (PD) games with real monetary stakes (*23*). Potential partners and ego-centric game-networks were visually presented on individual computer screens. Subjects played 20 rounds of the experimental game, in which a round consisted of a partner selection stage and a PD game stage. Two individuals were paired if and only if both proposed to each other at the partner selection stage (Fig. 1). Subjects played the 2-person PD game for each established link. A subject could differentiate his/her PD game strategies in games with different partners. Subjects paid link costs of $4.4(k-1)^2$ in the low-cost setting and $8.8(k-1)^2$ in the high-cost setting for $k>1$, and 0 otherwise, where $k$ denotes the number of established links. The link cost limited the number of links a subject could profitably maintain. We conducted four sessions in total, two for each cost setting.

<Figure 1 Here>

Cooperation evolved in all sessions, with high levels of cooperation in most rounds. The proportion of cooperative choices (%$C_{total}$) from round 1 to round 19 was 59.89 (±11.26)% in the low-cost setting and 76.37 (±9.66)% in the high-cost setting (Fig. 2A). The %$C_{total}$ in the first rounds (53.89±9.46%) were similar to those in the fixed-matching PD experiments (*24, 25*). In our experiments, however, cooperation gradually increased, in contrast to the decreasing patterns found in most fixed-matching experiments (*24, 25*). Only in the very final round did the %$C_{total}$ plunge to 21.21(±6.73)%, showing a strong end-game effect (*23*).

The high level of cooperation is associated with an increasing level of positive assortment, a well-known driver of the evolution of cooperation (*7, 8, 10, 21, 26-28*). To measure the degree of positive assortment, we compare the observed proportions of cooperation-cooperation pairs and defection-defection pairs with the proportions that would be obtained from random interactions. Fig. 2B shows that the observed proportions were larger than the null proportions and the difference grew over time. (For the calculation of null values, see (*23*)). Furthermore, individuals with higher rates of cooperation were more likely to interact with each other as indicated by the increasing Pearson correlation coefficient between paired players' %C (Fig. 2C).

<Figure 2 here>

The degree of modularity observed in our experiments was significantly lower than that of the random benchmark (*23, 29*) (Fig. 3A). The presence of positive assortment without a significant modular structure suggests core-periphery segregation, in which cooperators occupy a highly connected, cohesive core and defectors scatter on the sparse, intransitive periphery (*30*).

To quantify the location of cooperators and defectors in the core-periphery spectrum, we adopted the *k*-shell decomposition method (Fig. 3B), which outputs an integer index of coreness ($k_s$) for each node (*31*). The method takes into account not only the number of links a node has but also the number of links each of its neighbors has. For example, a node with $k_s$ =3 has at least three neighbors each of whom has at least three neighbors and so on. The coreness metric better elucidates the structural position of a node compared to ego-centric measures such as simple degree (*k*) or clustering coefficient which does not discount connections to peripheral nodes (*23*).

<Figure 3 here>

Cooperators formed a stable and dense cluster at the structural core of the network, whereas defectors loosely and unstably connected to this core. Pearson correlation coefficients between the %C and the $k_s$ of nodes steadily increased until near the end (Fig. 3C). Figures 3D and 3E provide graphical illustrations of the core-periphery segregation. The core-periphery segregation was robust across all four sessions (*23*).

The driving force behind the core-periphery segregation was the subjects' networking strategies (i.e., partner selection) rather than their PD game choices for given partners. The subjects' choices in PD games rarely changed between rounds (fig. S7 and table S4). The %C profile vectors of subjects in two adjacent rounds were significantly correlated with each other ($r = 0.85\pm0.09$). The core-periphery segregation resulted from the subjects' networking strategies: cooperators used Quit-for-Tat (*7, 9*), and defectors used Roving (*11*). Cooperators never left cooperative partners, but they unforgivingly cut relationships with defectors. Defectors, in contrast, frequently left current partners to search for new victims even when the partners were cooperators (Fig. 4B).

<Figure 4 here>

As a consequence of the networking strategies used by cooperators and defectors, the cooperative dyads continued to next rounds with high probabilities (95.91±5.89%), whereas mutual defection and mixed (CD) dyads were unlikely to continue (13.54±12.38% and 21.00±10.09%, respectively) (Fig. 4C). As rounds repeated, deeper cores with higher $k_s$ values were formed. At the same time, the probability of link maintenance at the core grew larger whereas the links at the periphery became increasingly unstable (Fig. 4C). Defectors had difficulty in connecting to others. Even though defectors made much more proposals than cooperators, they had much smaller number of proposals actually accepted as indicated by the increasing level of correlations between the number of accepted proposals and %C (fig. S9).

Cooperators enjoyed payoff-advantages. In earlier rounds, defectors gained more than cooperators. Over time, however, cooperators' earnings grew larger than defectors' (Fig. 4D). Cooperators had large numbers of mutually cooperative links whereas defectors had fewer links and their links mostly ended up with mutual defection outcomes. This result suggests that the experimental conditions of partner choice and exit served as a mechanism for the evolution of cooperation. The proportion of cooperation increased as some players' cooperative behavior created incentives for others to behave likewise (*15*). But the early defectors often failed to join the cooperative core when they tried to convert in a later time because many cooperators were already locked in with each other.

Combining behavioral experiments and social network analysis, we examined the generative mechanism of dense cooperative clusters as the structural correspondence of social capital. Dense networks of cooperators emerged as a consequence of preferential partner choice even without mechanisms for social reputation or indirect reciprocity. Cooperators organized a closed network of their own, stably occupying the structural core of the evolving network. This result supplements the concept of spatial (*27*) and network reciprocity (*28*) by specifying the topological characteristics of type clusters and their locations in the core-periphery spectrum of the global network. The structural stratification produced inequality by providing a large material advantage to players located at the cooperative core and, thus, suppressed the revival of defectors. In other words, the collective pattern of network formation served as a self-organized punishment mechanism toward defection.

Our experiment shows the feasibility and characteristics of self-organized network solutions for the dilemma of cooperation. Our experiment raises several questions for future research. If endogenous network formation is a common characteristic of contemporary human interactions, why do some communities, groups, and societies fail to evolve dense cooperative clusters? Under what conditions could the defectors at the periphery change their strategies and join the cooperators' cluster? Research on these questions will provide a more comprehensive picture of the interplay between strategic choices and network structures.

**References and Notes**

1. S. Bowles, H. Gintis, Social Capital And Community Governance. *The Economic Journal* **112**, F419 (2002).
2. E. Ostrom, T. K. Ahn, Eds., *Foundations of social capital*, (Edward Elgar, Cheltenham, UK, 2003).
3. R. D. Putnam, R. Leonardi, R. Nanetti, *Making democracy work : civic traditions in modern Italy*. (Princeton University Press, Princeton, N.J., 1993).
4. J. Bascompte, P. Jordano, J. M. Olesen, Asymmetric coevolutionary networks facilitate biodiversity maintenance. *Science* **312**, 431 (2006).
5. B. Kerr, C. Neuhauser, B. J. M. Bohannan, A. M. Dean, Local migration promotes competitive restraint in a host–pathogen'tragedy of the commons'. *Nature* **442**, 75 (2006).
6. G. Kossinets, D. J. Watts, Empirical analysis of an evolving social network. *Science* **311**, 88 (2006).
7. N. Hayashi, T. Yamagishi, Selective play: Choosing partners in an uncertain world. *Personality and Social Psychology Review* **2**, 276 (1998).
8. M. W. Macy, J. Skvoretz, The evolution of trust and cooperation between strangers: A computational model. *American Sociological Review*, 638 (1998).
9. S. Moresi, S. Salop, in *Economics for an imperfect world: essays in honor of Joseph E. Stiglitz,* R. Arnott, B. Greenwald, R. Kanbur, B. Nalebuff, Eds. (MIT Press, 2003).
10. F. C. Santos, J. M. Pacheco, T. Lenaerts, Cooperation prevails when individuals adjust their social ties. *PLOS Computational Biology* **2**, e140 (2006).
11. L. A. Dugatkin, D. S. Wilson, Rover: a strategy for exploiting cooperators in a patchy environment. *American Naturalist*, 687 (1991).
12. R. Axelrod, W. Hamilton, The evolution of cooperation. *Science* **211**, 1390 (1981).
13. M. A. Nowak, Five rules for the evolution of cooperation. *Science* **314**, 1560 (2006).
14. R. L. Trivers, The evolution of reciprocal altruism. *Quarterly Review of Biology*, 35 (1971).
15. C. F. Camerer, E. Fehr, When does" economic man" dominate social behavior? *Science* **311**, 47 (2006).
16. P. Hammerstein, in *Genetic and Cultural Evolution of Cooperation,* P. Hammerstein, Ed. (MIT Press, 2003).
17. J. Orbell, R. M. Dawes, A" cognitive miser" theory of cooperators' advantage. *American Political Science Review*, 515 (1991).
18. E. L. Thorndike, D. Bruce, *Animal intelligence: Experimental studies*. (Macmillan, New York, 1911).
19. G. Tullock, Adam Smith and the prisoners' dilemma. *The Quarterly Journal of Economics*, 1073 (1985).
20. A. L. Barabási, R. Albert, Emergence of scaling in random networks. *science* **286**, 509 (1999).
21. B. Skyrms, R. Pemantle, A dynamic model of social network formation. *Proceedings of the National Academy of Sciences* **97**, 9340 (2000).
22. K. Fehl, D. J. van der Post, D. Semmann, Co-evolution of behaviour and social network structure promotes human cooperation. *Ecology Letters* **14**, 546 (2011).
23. See supporting material.
24. J. Andreoni, J. H. Miller, Rational cooperation in the finitely repeated prisoner's dilemma: Experimental evidence. *The Economic Journal*, 570 (1993).

25. R. Cooper, D. V. DeJong, R. Forsythe, T. W. Ross, Cooperation without reputation: experimental evidence from prisoner's dilemma games. *Games and Economic Behavior* **12**, 187 (1996).
26. C. L. Apicella, F. W. Marlowe, J. H. Fowler, N. A. Christakis, Social networks and cooperation in hunter-gatherers. *Nature* **481**, 497 (2012).
27. M. A. Nowak, R. M. May, Evolutionary games and spatial chaos. *Nature* **359**, 826 (1992).
28. H. Ohtsuki, C. Hauert, E. Lieberman, M. A. Nowak, A simple rule for the evolution of cooperation on graphs and social networks. *Nature* **441**, 502 (2006).
29. M. Newman, Communities, modules and large-scale structure in networks. *Nature Physics* **8**, 25 (2011).
30. S. P. Borgatti, M. G. Everett, Models of core/periphery structures. *Social Networks* **21**, 375 (2000).
31. M. Kitsak *et al.*, Identification of influential spreaders in complex networks. *Nature Physics* **6**, 888 (2010).

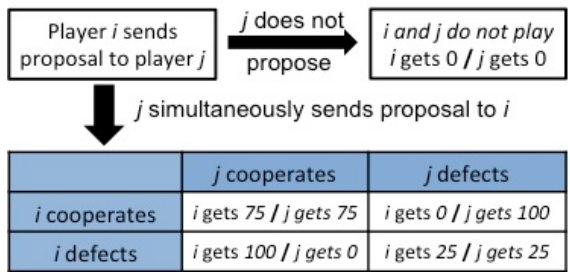

**Fig. 1.** The PD game with partner choice and exit options. Two subjects play the PD game only when they both propose to play with each other at the partner selection stage of a given round. After a link is formed, each dyadic pair plays the PD game.

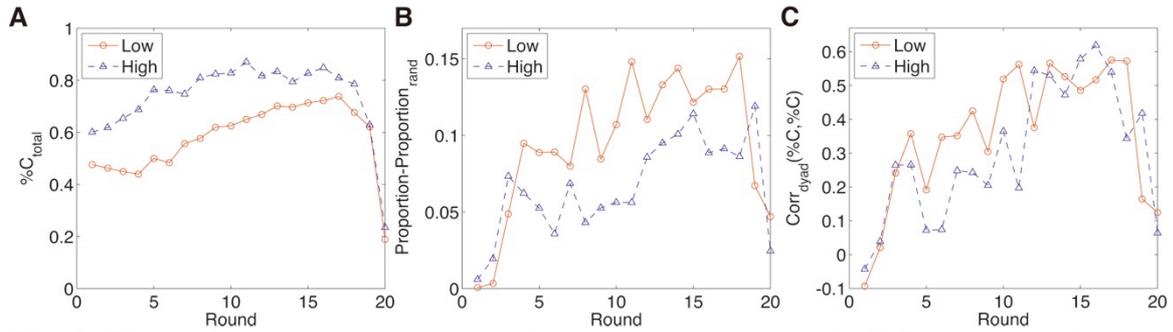

**Fig. 2.** The evolution of cooperation and positive assortment. (**A**) %C$_{total}$ increased over time. Each point on the lines represents the proportion of cooperative choices, calculated by the number of cooperative choices divided by the number of all choices made by all subjects in a given round. (**B**) The proportion of CC (or DD) minus the proportion obtained by random shuffling of the game strategies grew over time (*23*). (**C**) The Pearson correlation coefficient between paired players' %C increased over time until near the final round in both cost settings.

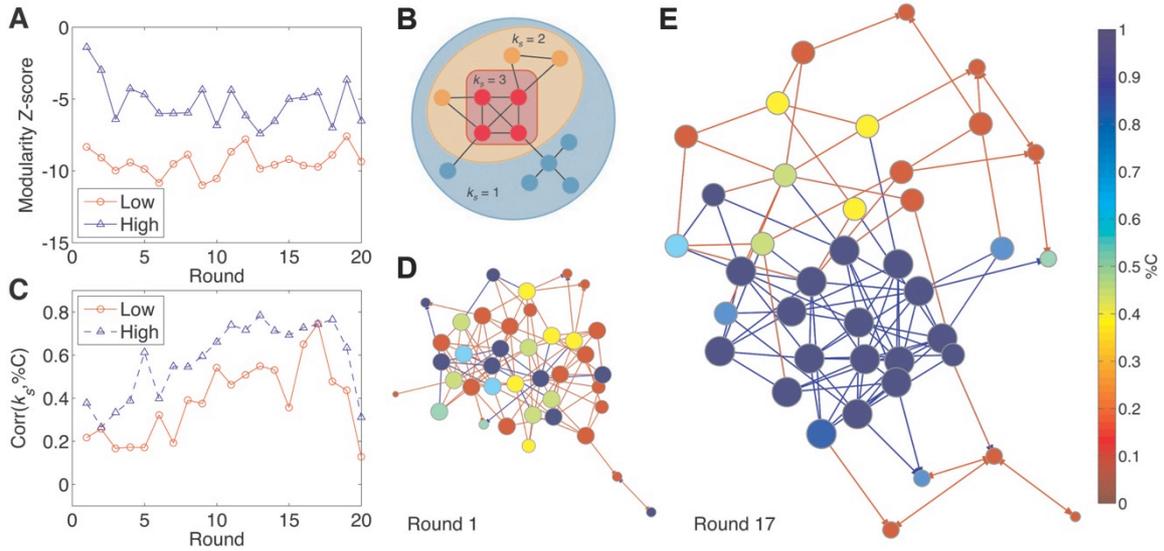

**Fig. 3.** Core-periphery segregation. (**A**) Z-scores of modularity value of each interaction network per round compared to one thousand random networks preserving the degree distribution of a given round (*23*). (**B**) A toy network consisting of nodes with different coreness ($k_s$). Nodes are assigned with $k_s$ indicating whether they are located at the densely connected core or the sparsely connected periphery (*31*). The iterative pruning process begins by eliminating all nodes with degree $k = 0$. This process is repeated until there is no node left with $k = 0$. Nodes pruned in this round are assigned the $k$-shell layer $k_s = 0$. Next, remaining nodes with $k \leq 1$ were pruned and assigned with $k_s = 1$. This process was repeated by sequentially (+1) increasing the number of maximum degree ($k$) and $k_s$ until all of the nodes were removed. (**C**) Correlation between a player's coreness ($k_s$) and %C. High cooperators were more likely to be located at the core as rounds repeated. (**D**) and (**E**) Graphical illustration of core-periphery segregation in two representative rounds of a low-cost session. The color of each node indicates %C of the depicted player in the round. A blue arc with arrow indicates its originator's cooperative decision to its receiver, whereas an arc in red represents defection. Nodes are distributed in a spring-embedded layout that locates transitively connected nodes at proximal distances. The larger node indicates higher $k_s$.

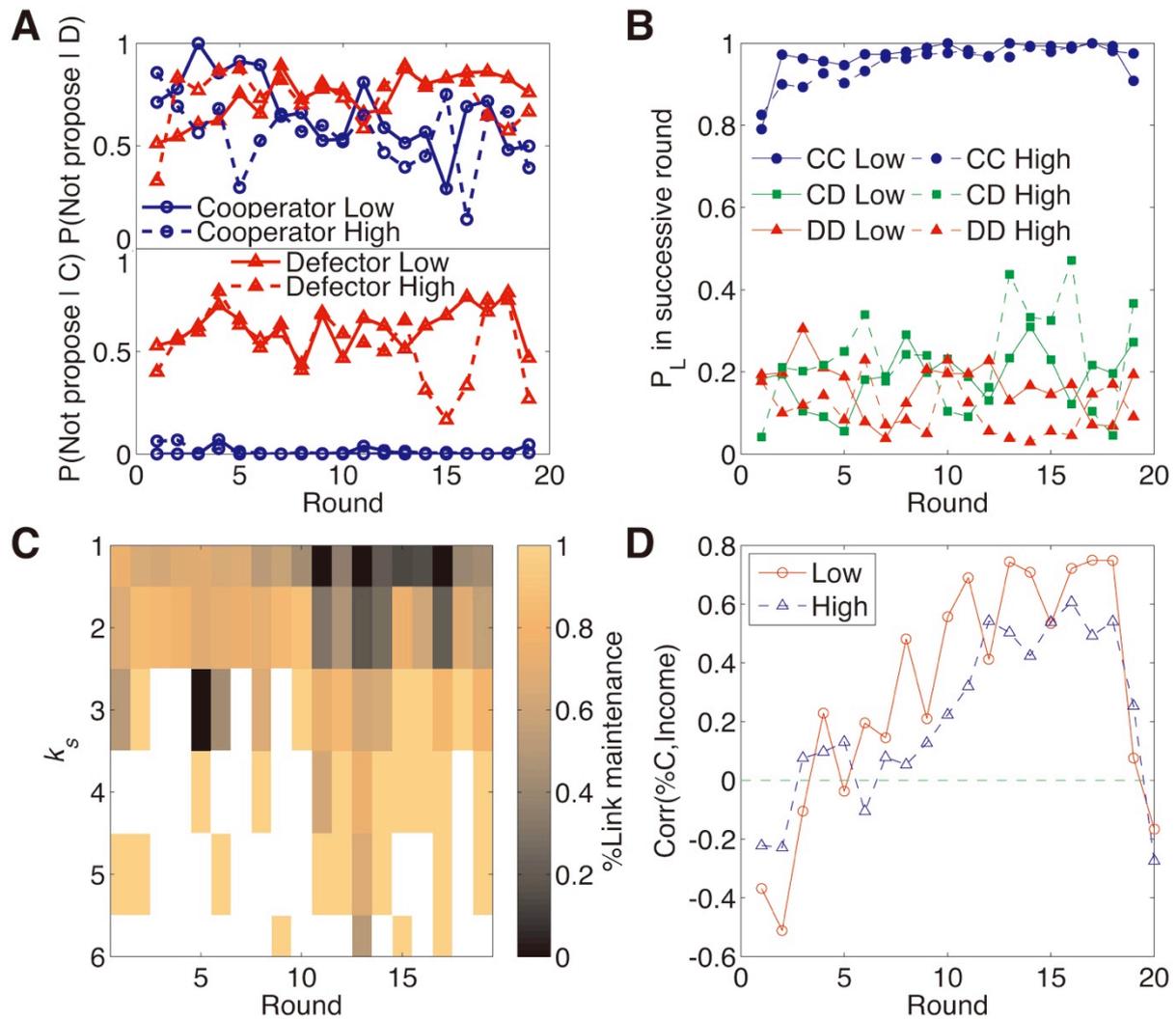

**Fig. 4.** The coevolution of network and strategy. (**A**) The top panel shows the probabilities of not proposing to defectors of the current round in the next round. The bottom panel shows the probabilities of not proposing to cooperators of the current round in the next round. Blue and red colors correspond to the ego game strategies: 100% cooperators and 100% defectors, respectively. (**B**) The probability of link continuation as a function of the outcome in the preceding round. Subjects regardless of their own types cut links with defectors (top panel). Cooperators tried to keep the links with cooperators whereas defectors hit and ran (bottom panel). (**C**) The probability of link continuation to the next round given the players' coreness measured in the current round. The first session of the low-cost setting is illustrated. See fig. S8 for full results. (**D**) The tendency of more cooperative subjects receiving higher earnings increased over time.

# Supplementary Information for

# Core-Periphery Segregation in Evolving Prisoner's Dilemma Networks

Yunkyu Sohn

Jung-Kyoo Choi

T.K. Ahn

December 8, 2012

# Contents

# 1 Experimental Procedure with Sample Screen Shots

**Experimental procedure**. The experiments were conducted in a computer lab at Korea University, Seoul, Korea in June and July of 2010. Paid volunteer subjects were recruited via web advertisements posted on university websites. A total of 150 subjects participated in the experiments, 35 and 40 in two high-cost sessions and 36 and 39 in two low-cost sessions. The server-client type experimental program was written in C#. Dividers were set between monitors to provide privacy during decision making. Subjects were informed that there would be 20 fixed rounds. Subjects saw others as scattered nodes on their computer screen and could not identify which node represented which other person in the room. Several sample screen shots are provided in the Supplementary Information. Subjects accumulated earnings in points during the experiment. An initial endowment of 500 points was given at the beginning of the experiment. One point was converted to 4 Korean Won at the conclusion of the experiment and paid in cash, in addition to a 5,000 Won show-up fee. On average, subjects earned 19,414 Won, with a maximum of 32,244 Won and a minimum of 5,000 Won, including the show-up fee. (the exchange rate was 1,214 Won for 1 US dollar at the time of the experiment).

One round of the experiment consisted of the following steps:

**Proposal Stage.** On the individual computer screen, each subject's own node was marked black, while other nodes were hollow circles (fig. S1). Subjects proposed to others to play the Prisoner's Dilemma (PD) game by clicking up to twenty other nodes. The chosen nodes turned green. Each subject did not know who, if anyone, was going to propose to play the game with her/him.

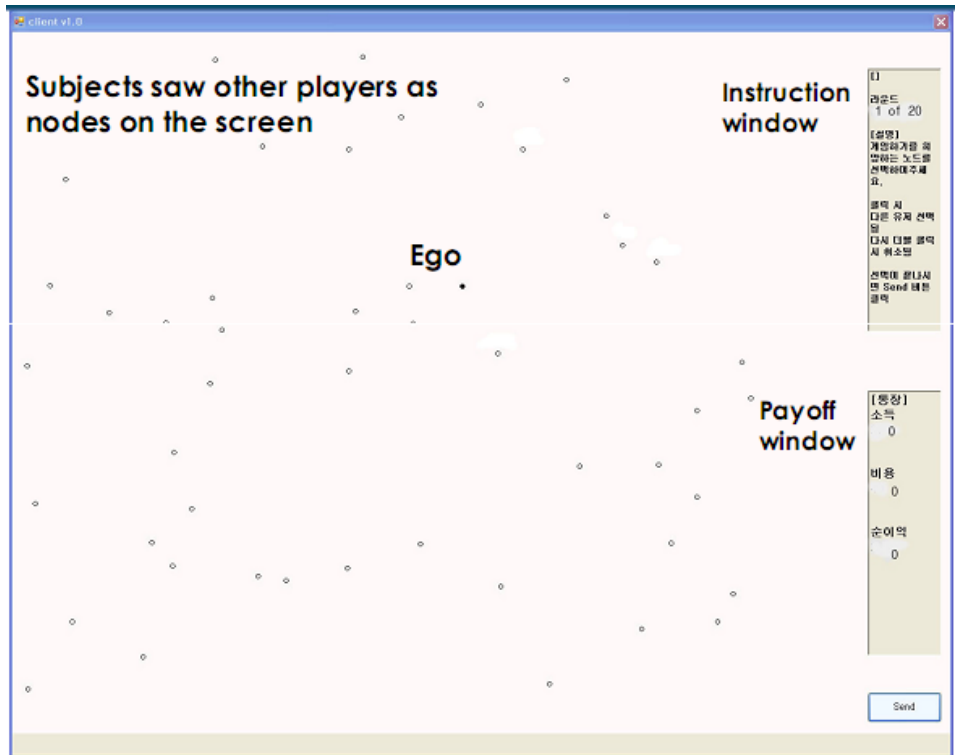

**Figure S1.** Proposal stage sample screen.

**PD Game Stage.** When two subjects proposed to each other in the proposal stage of a round, a link was shown connecting the two nodes at the beginning of the PD game stage (fig. S2). For each connected node, subjects chose their game strategy by clicking the node that turned the node red (cooperation) or blue (defection). On the PD game stage screen, subjects' past interaction history was provided for up to the past three rounds. This information was provided to each of the other subjects with whom a link was established at least one time during the last three rounds. The subjects' history information was shown by a string of three symbols next to alters' nodes, representing cooperation (by the alter), defection (by the alter), and no link (fig. S3). For example, A0B next to an alter's node meant that the alter cooperated three rounds prior, was not linked to ego two rounds prior, and defected in the immediately preceding round.

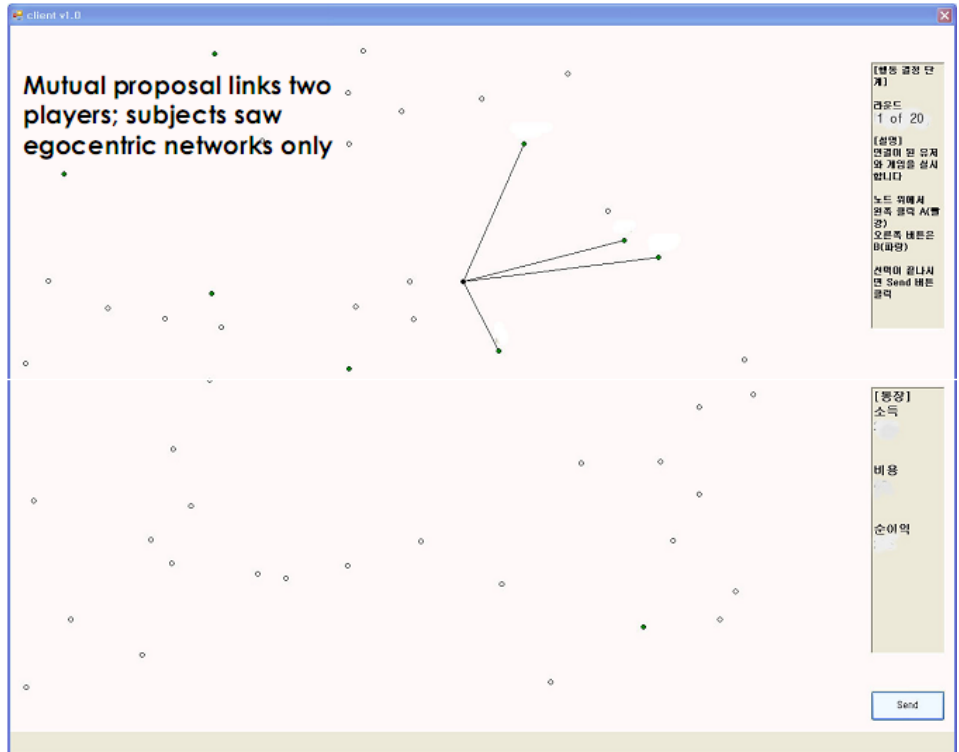

**Figure S2.** PD game stage sample screen.

**Round Outcome Feedback.** After all subjects made game decisions, game outcomes were displayed graphically on the feedback screen by the colors of the links. Red indicated cooperation and blue indicated defection. The color of the half of the link that originated from the ego represented the ego's strategy choice, and the color of the half of the link that originated from the alter represented the alter's strategy choice (fig. S4). On the feedback screen for each round, the ego's total earning in that round from all of the games she played, the total cost as a function of the number of links in that round, and the net earning in that round were shown in the payoff window at the bottom right corner of the computer screen.

After Round 1, if a pair of subjects were linked, which meant they played the PD game in the immediately preceding round, they saw a black line connecting their nodes on the proposal stage screen. However, if either subject did not propose to the other in the proposal stage, the line disappeared on the screen at the PD game stage.

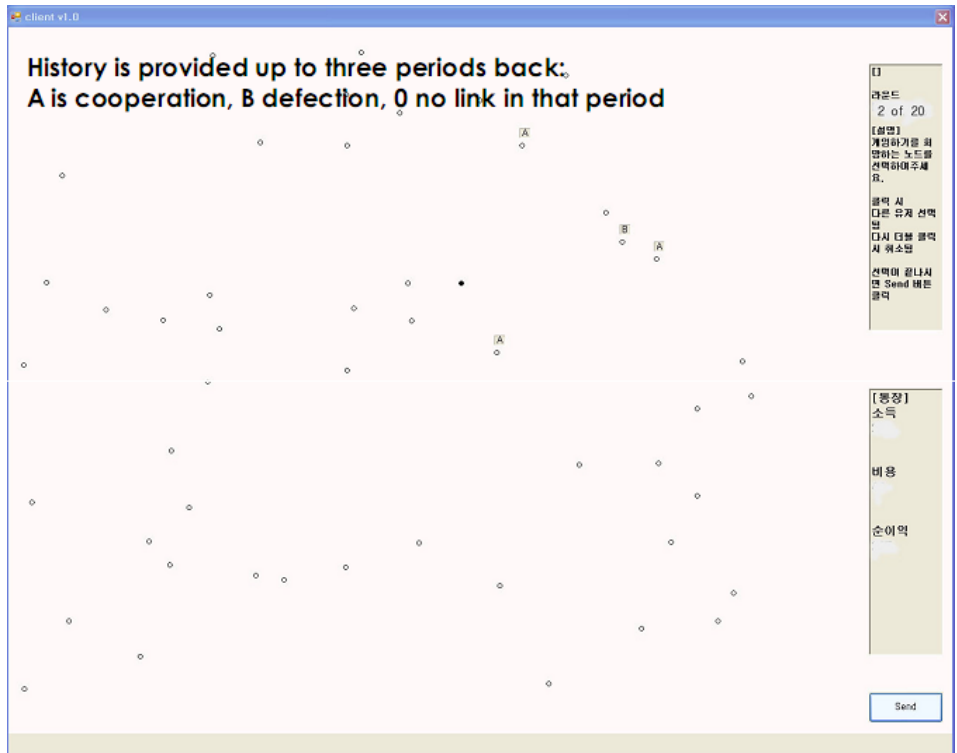

**Figure S3.** Providing interaction history.

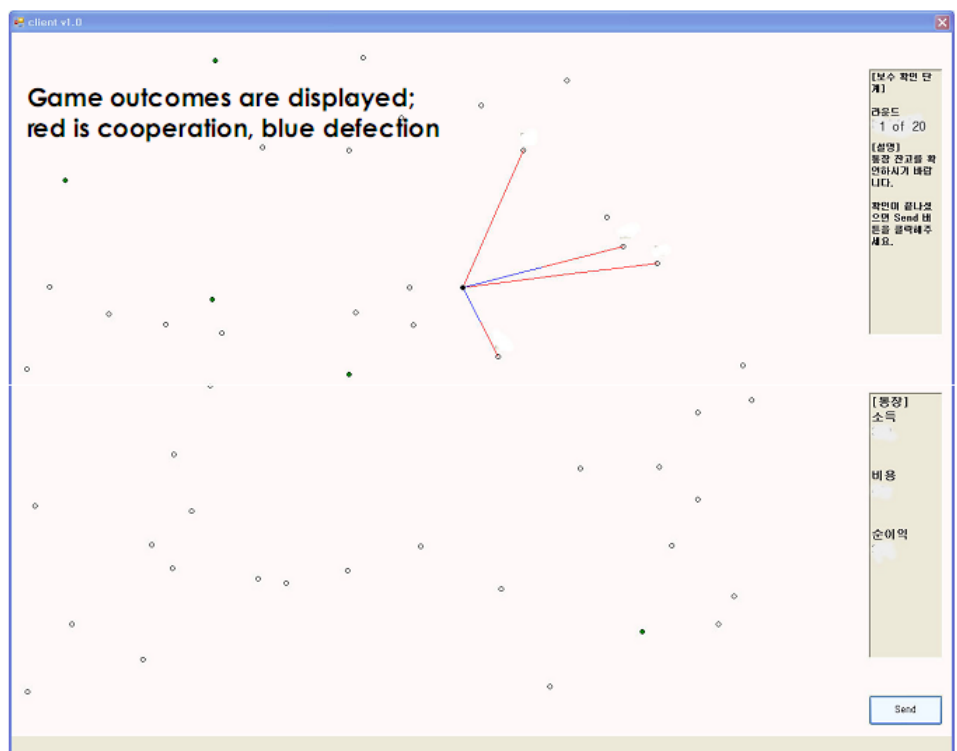

**Figure S4.** Display of game outcomes.

## 2 METHODS

**Randomization procedure for Fig. 2B** To obtain the null values, game strategies (C and D) were randomly redistributed over the nodes while fixing the structure of the network and the number of Cs and Ds in the population. This procedure was performed for each round in each session. The values shown in the figure are the averages of 1,000 iterations of the procedure.

**Computation of Modularity Z-score via link swapping** To characterize the modular structure in each network, we utilized a modularity-based community detection method. The benefit function, modularity $Q$, is defined as follows (*29, 32, 33*):

$$Q = \sum_{i=1}^{N_M} \left[ \frac{l_s}{L} - \left( \frac{d_s}{2L} \right)^2 \right] \qquad (1)$$

where $N_M$ is the number of modules, $l_s$ is the number of links within module $s$, $L$ is the total number of links in the network and $d_s$ is the total number of links that are connected with nodes belonging to module $s$. We employed an iterative greedy algorithm to find an optimal division corresponding to the maximum modularity value ($Q_{max}$) of the interaction network for a given round (*32*). The corresponding modularity value ($Q_{max}$) for that particular division was used to indicate the level of natural clustering in the network. (See S4.3.1 for details.)

To obtain random reference networks for each round's interaction network while preserving its degree distribution, we used link swapping method (*34*). For each step of link swapping, two links with non-overlapping nodes were selected. Suppose the first link was a connection between nodes a and b and the second link was a connection between nodes c and d. Then, these links swapped their neighbors so that after the swapping, links a-b and c-d were eliminated and new links a-d and c-b were formed. We repeated this procedure at least 5 times for each link to obtain each of the 1,000 random networks for a given round.

Finally, modularity Z-score for a given round was computed as follows:

$$Q_{Z-score} = \frac{Q_{max} - \overline{Q_{max}^{rand}}}{\sigma(Q_{max}^{rand})}$$

(2)

where $Q_{max}$ is the maximum modularity value of the optimal division for the network of a given round, $\overline{Q_{max}^{rand}}$ is the average and $\sigma(Q_{max}^{rand})$ is the standard deviation of maximum modularity values of 1,000 random reference networks respectively.

**The *k*-shell decomposition method.** In each round, coreness values ($k_s$) were assigned to nodes according to whether they were located at the densely connected core or the sparsely connected periphery (Fig. 3B) (*31, 35, 36*). The iterative pruning process began by eliminating all nodes with degree $k = 0$. This process was repeated until there was no node left with $k = 0$. Nodes pruned in this round were assigned the *k*-shell layer $k_s = 0$. Next, remaining nodes with $k \leq 1$ were pruned and assigned with $k_s = 1$. This process was repeated by sequentially (+1) increasing the number of maximum degree ($k$) and $k_s$ until all of the nodes were removed.

## 3 Individual Incentives in the Experimental Setting

### 3.1 Optimal Number of Links as a Function of Link Costs and Game Outcomes

The link cost function and the game outcomes jointly determine the optimal number of links that an earnings-maximising player would wish to establish. If the game is played for only a single round, a rational and self-interested player would defect in all PD games that he plays in the PD game stage. If all players are rationally self-interested and this is common knowledge, then all players would expect everyone else to defect in the PD games. If a player plays $k$ games in a round, then the total payoff from the PD games is $25k$, as the per game payoff from the PD game is 25 (Fig. 1). With this expectation of the game outcomes, a player's expected net earnings in a round, as a function of the number of established links (i.e., the games actually played), is $25k - a(k-1)^2$, where $a = 4.4$ in the low cost setting and 8.8 in the high cost setting. The integer value of $k$ that maximizes this payoff function is 4 in the low cost setting and 2 in the high cost setting.

When the game is repeated, or if the players are not strict payoff maximizers, the expectation for the game outcomes in a round may differ from the universal defection. Suppose that in a given round, a player expects all of the others with whom he plays the PD game to cooperate, and thus he himself plans to cooperate because he is a conditional cooperator, or because he expects future gains from current cooperation. In this situation, the total payoff from the PD game is $75k$, where $k$ is the number of games and 75 is the payoff corresponding to the CC (mutual cooperation) outcome. The expected payoff as a function of the number of links is $75k - a(k-1)^2$, and the integer $k$ value that maximizes this function is 10 in the low cost setting where $a = 4.4$ and 5 in the high cost setting where $a = 8.8$.

Note that in the low cost setting, the net earning is 394 when $k = 10$ and all outcomes are mutual cooperation (CC), which is the maximum that cooperators can achieve. The experimental results show that the number of links that subjects build seldom reaches the optimal number of 10. The reason for this is because the net earning is already 340 when $k = 6$ and all outcomes are CC. Thus, the marginal gain from additional CC links is small relative to the risks that an intending cooperator must take when encountering defectors and searching for more cooperators. If a new partner defects, the large marginal link cost at that level of links becomes a pure net cost because the payoff from the PD game is zero when the outcome is CD.

The optimal number of links when expected PD game outcomes are all CD or DC can be similarly calculated. When a player expects mixed game outcomes, the optimal number of links depends on the distribution of game outcomes. For example, if a player intends to cooperate in every game and expects 50% of the partners to cooperate, then the per-game expected payoff is 50, and the optimal number of links is 7 in the low cost setting and 4 in the high cost setting. Of course, if a player knows precisely who are the defectors and cooperators, he would link to those players who guarantee him larger payoffs and would add other partners as long as the marginal net gain stayed above zero.

In a dynamic context in which a rational player tries to maximize the total earning in the repeated game (20 rounds in our experiments), he might be willing to incur a net cost in the current round for future gains. This is the case, for example, when a conditional cooperator wishes to quickly build up as many mutually cooperative links as possible, up to a desired number.

**3.2 Strategies for the PD games with Exit and Partner Choice Options**

A player's strategy for this game consists of two inter-related sub-strategies. The PD game strategy refers to plans to either cooperate or defect with each of the connected partners. The networking strategy refers to the decision rule about to whom to propose and whether to continue or cut a current link. The strategy that maximizes one's monetary payoff depends on how others play the game. If most others intend to cooperate and cut links with defectors, a strategy referred to as Out-for-Tat (*7, 37*) or Quit-for-Tat (*9*), then the best way to maximize ones monetary income is to cooperate until it is close to the end of the repeated game. If there are many cooperators and they use the Quit-for-Tat strategy, roving defectors may not succeed in exploiting cooperators. However, if there are too few cooperators, then trying to accumulate cooperative relationships would incur a large search cost due to random encounters with defectors. In this case, a player would limit the search and be satisfied with a few links with suboptimal game outcomes. An optimist who believes that most others would play the game in a conditionally cooperative manner would also start the game cooperatively. A pessimist who mistrusts others, or who is too cautious to expose himself in early rounds, even if his intrinsic preference is conditionally cooperative, would build a limited number of relationships and would defect in games.

## 4 Additional Analyses of the Experimental Data

### 4.1 Cooperation and Earnings

#### 4.1.1 Cooperation

The proportion of cooperation (%$C_{total}$) in a round is defined as the proportion of the number of A choices (cooperation) over the number of the entire games played in the reference group. For example, %$C_{total}$ in Round 1 of low cost setting is the number of A choices played in Round 1 in the low cost setting in proportion to the total number of games played in Round 1 in the low cost setting. Thus, if a player played four games and cooperated in three of them, and another player played six games and cooperated in one of them, the %$C_{total}$ in this group of two players is $(4/10) \times 100\% = 40\%$. This is different from averaging the %C's between the two players, which would be 45.9%. The formula we used is %$C_{total} = \sum_{i,j} A^C_{ij} / (\sum_{i,j} A^C_{ij} + \sum_{i,j} A^D_{ij})$, where $A^C$ and $A^D$ are asymmetric $N \times N$ matrices, illustrating cooperative and non-cooperative strategic profiles, respectively, among $N$ subjects in a given round.

**Table S1.** The rates of cooperation by session.

| Round | Low Cost Setting | | High Cost Setting | |
|---|---|---|---|---|
| | Session 1 | Session 2 | Session 1 | Session 2 |
| 1 | 0.50 | 0.45 | 0.53 | 0.67 |
| 2 | 0.44 | 0.49 | 0.53 | 0.71 |
| 3 | 0.39 | 0.51 | 0.56 | 0.74 |
| 4 | 0.37 | 0.51 | 0.63 | 0.75 |
| 5 | 0.46 | 0.54 | 0.76 | 0.77 |
| 6 | 0.42 | 0.54 | 0.78 | 0.74 |
| 7 | 0.46 | 0.66 | 0.73 | 0.76 |
| 8 | 0.51 | 0.64 | 0.80 | 0.82 |
| 9 | 0.54 | 0.70 | 0.78 | 0.87 |
| 10 | 0.59 | 0.66 | 0.79 | 0.86 |
| 11 | 0.64 | 0.66 | 0.87 | 0.87 |
| 12 | 0.61 | 0.72 | 0.77 | 0.86 |
| 13 | 0.72 | 0.68 | 0.85 | 0.81 |
| 14 | 0.68 | 0.71 | 0.72 | 0.87 |
| 15 | 0.71 | 0.71 | 0.79 | 0.87 |
| 16 | 0.68 | 0.76 | 0.88 | 0.82 |
| 17 | 0.73 | 0.75 | 0.83 | 0.79 |
| 18 | 0.62 | 0.74 | 0.77 | 0.80 |
| 19 | 0.63 | 0.61 | 0.70 | 0.56 |
| 20 | 0.26 | 0.11 | 0.25 | 0.22 |

Our experimental results are distinct compared to other repeated PD experiments in that a general tendency of increasing rates of cooperation was observed in all sessions, although some fluctuation and decline at the end was observed. The exact values for each session are provided in table S1. The rates of cooperation were higher in the high cost setting than the low cost setting. In addition, the cooperation rates approached the peak values more quickly in the high cost setting. It appears that the smaller number of optimal CC links for cooperators in the high cost setting, compared to the low cost setting (5 vs. 10), allowed the subjects in the high cost setting to be optimistic about starting the game cooperatively and to thus quickly build close-to-optimal numbers of CC links. The high rates of cooperation were sustained until Round 19. The exact point from which the rates generally decreased varied across sessions. The final round outcome was close to the single-round game benchmarks (see S3.1 for details).

**4.1.2 Earnings**

There are two benchmarks against which to compare the actual earnings that the experimental subjects achieved. The first is the single-round game efficient subgame-perfect equilibrium payoff. This is achieved when each player has four (in the low cost setting) or two (in the high cost setting) links and defects in all games. The per-period net earning in this case is 60 in the low cost setting and 41 in the high cost setting (see S3.1 for details). The other benchmark is the social optimum, which involves everyone cooperating in every game. The social optimum requires that everyone has exactly 10 (in the low cost setting) or 5 (in the high cost setting) games and cooperates in all games. In this case, the net earning for each player is 394 in the low cost setting and 234 in the high cost setting. The fact that our subjects achieved earnings that were much higher than the single-round rational benchmark is implied in the cooperation rates. However, the exact earning was not just a function of the rates of cooperation, but was also a function of the number of games they played. In addition, there was substantial variability in earnings between subjects, as implied by core-periphery segregation of players.

The average per-round earnings were well-above the single-round efficient equilibrium benchmarks in both settings: 166.5 and 186.4 (compared to a benchmark of 60) in the two low cost sessions and 119.3 and 129.5 (compared to a benchmark of 41) in the high cost sessions. The per-round averages followed the trends in the rates of cooperation (table S2). However, these averages were much smaller than the social optimum payoffs. There was substantial variability across individuals in total net earnings for the entire 20 rounds (table S3).

**Table S2.** Subjects' average net earnings per round by session.

|   | Low Cost | | High Cost | |
|---|---|---|---|---|
|   | Session 1 | Session 2 | Session 1 | Session 2 |
| 1 | 143.2 | 112.3 | 68.9 | 94.5 |
| 2 | 133.2 | 156.1 | 66.6 | 122.8 |
| 3 | 125.9 | 162.7 | 64.6 | 122.9 |
| 4 | 127.3 | 168.6 | 114.9 | 108.7 |
| 5 | 129.3 | 175.8 | 122.2 | 107.4 |
| 6 | 135.3 | 164.1 | 118.6 | 104.9 |
| 7 | 147.9 | 182.0 | 129.0 | 131.2 |
| 8 | 154.3 | 196.3 | 135.7 | 143.2 |
| 9 | 163.1 | 212.3 | 124.2 | 148.5 |
| 10 | 174.6 | 210.3 | 126.1 | 155.2 |
| 11 | 185.4 | 206.8 | 141.1 | 149.4 |
| 12 | 190.7 | 209.8 | 145.6 | 165.4 |
| 13 | 189.1 | 216.3 | 140.9 | 155.6 |
| 14 | 212.2 | 223.8 | 111.4 | 161.3 |
| 15 | 216.5 | 221.1 | 134.3 | 157.0 |
| 16 | 211.0 | 231.1 | 161.5 | 154.2 |
| 17 | 215.9 | 230.7 | 146.0 | 148.4 |
| 18 | 200.7 | 212.5 | 152.5 | 132.6 |
| 19 | 172.7 | 180.5 | 131.5 | 82.8 |
| 20 | 101.3 | 54.2 | 50.1 | 43.8 |

**Table S3.** Total earnings for the entire 20 rounds. (There were three subjects whose total earnings fell below zero. These subjects were paid only the show-up fee of 5,000 Won.)

|       | Low Cost   |           | High Cost  |           |
|-------|------------|-----------|------------|-----------|
|       | Session 1  | Session 2 | Session 1  | Session 2 |
| N     | 39         | 36        | 35         | 40        |
| Mean  | 3330       | 3727      | 2386       | 2590      |
| Std   | 1064       | 1554      | 1265       | 1502      |
| Min   | 1854       | 1725      | -1317      | -4727     |
| 25th  | 2522       | 2449      | 1878       | 1698      |
| 75th  | 4119       | 5142      | 3328       | 3508      |
| Max   | 5872       | 6311      | 4237       | 4222      |

## 4.2 Positive Assortment

The emergence of positive assortment is captured in Fig. 2B and C. Figure 2B shows that the Pearson correlation coefficients of the rates of cooperation between dyadic pairs increased over rounds. Figure 2B reveals a statistically significant over-representation of both types of homophilous associations (CC and DD links) (See S2 Methods for details). The level of over-representation increased over time.

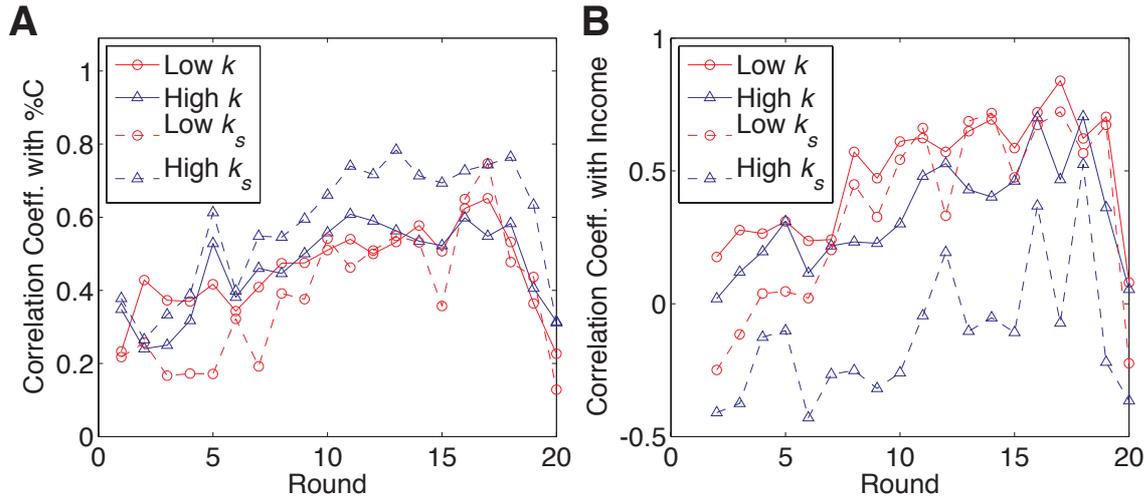

**Figure S5.** Additional analysis of the coevolution of network and strategy. All lines in the graphs are averages over two sessions for each cost setting. (**A**) Correlation coefficients between coreness ($k_s$) (dashed) and degree ($k$) (solid) and %C of nodes in each round. (**B**) Correlation coefficients between coreness ($k_s$) (dashed) and degree ($k$) (solid) and income gain of nodes in each round.

### 4.3 Core-periphery Segregation

#### 4.3.1 Modular Networks vs. Core-periphery Organizations

Core-periphery structure is a characteristic term that illuminates the global organization of a network. Although the community detection method (see S2 Methods for details), which is one of the most popular methods for analysing the global organization of a network (*29*), implicitly assumes the existence of a distinct modular structure in a network of interest, many empirical networks do not exhibit such a trait (*33, 38*). For a network to be characterized as having a modular structure, it should have dense modules that are connected to each other with marginal connections. Otherwise, the existing community detection methods end up detecting non-significant modules, despite the fact that they maximize the given benefit function (e.g., modularity) for module detection defined over the global network topology (*33*).

Many real and artificial networks are better illustrated using the core-periphery perspective (*30, 31, 35, 36, 39-42*). Core-periphery structure consists of dense, cohesive cores and sparse, intransitive peripheries. The former exhibits significantly transitive intra-connections, whereas the latter does not show such a level of cohesion. In other words, in contrast to a modular structure that consists of multiple clusters with a comparable status, core-periphery structure involves hierarchy and asymmetry. Let us illustrate the consequence of such a difference in terms of the highest possible values of the benefit function (e.g., modularity) for module detection of these two types of networks. Imagine that networks A and B have the same degree distributions, as well as the same size (# nodes: $N$) and density. It is possible that network A is better characterized as a modular network and network B is better characterized as a core-periphery organization. This is the case if network A consists of marginal connections between highly dense and transitive associations, whereas network B entails a few exceptionally dense, transitive associations that are connected to multiple sparse, intransitive associations. For simplicity, let us also assume that these two networks have identical association or module assignments. In other words, while these two networks have identical module assignment vectors (e.g., ($m_1$, $m_2$, $m_2$, $m_3$, $m_3$ ...), in which $m_i$ indicates that the corresponding node's module assignment is an $N \times 1$ vector), the link allocations over networks A and B are distinct. Network A entails multiple dense modules (e.g., intra-module density: 0.5) that are connected to each other with sparse connections, whereas network B entails a few exceptionally well-connected modules (core; e.g., intra-module density: 0.8) that are tied with low density modules (periphery; e.g., intra-module density: 0.1). Due to this difference, the highest possible values of the benefit function (e.g., modularity) for module detection of these two networks should exhibit different levels. Asymmetry in intra-density of the core and the periphery results in a lower maximum modularity value of network B than that of network A.

Maximum modularity, $Q_{max}$, (see Methods) captures the level of natural compartmentalisation embedded in the network of interest. Given that these two networks have the same degree distributions, and assuming that they have identical module assignment corresponding to $Q_{max}$ of each network, then the summed values of the latter term in equation (1), which represents the null expected values of link formation, are identical in networks A and B for every module. However, by definition, the summed values corresponding to the former term of each module, which characterizes the summation of intra-module density, should be lower in network B than in network A. Consequently $Q_{max}$ of network A, the modular network, is higher than $Q_{max}$ of network B, the core-periphery network. This observation suggests an instrument to differentiate core-periphery organizations from modular networks.

We quantitatively assess above predictions by looking at the properties of synthetic networks. Figure S6 clearly demonstrates that the original networks with core-periphery structure turn into significantly modular networks when randomized while preserving degree distributions. Over the wide range of network density, modularity values of the core-periphery organizations exhibit substantially lower values than those of the randomized networks confirming the plausibility of the above argument.

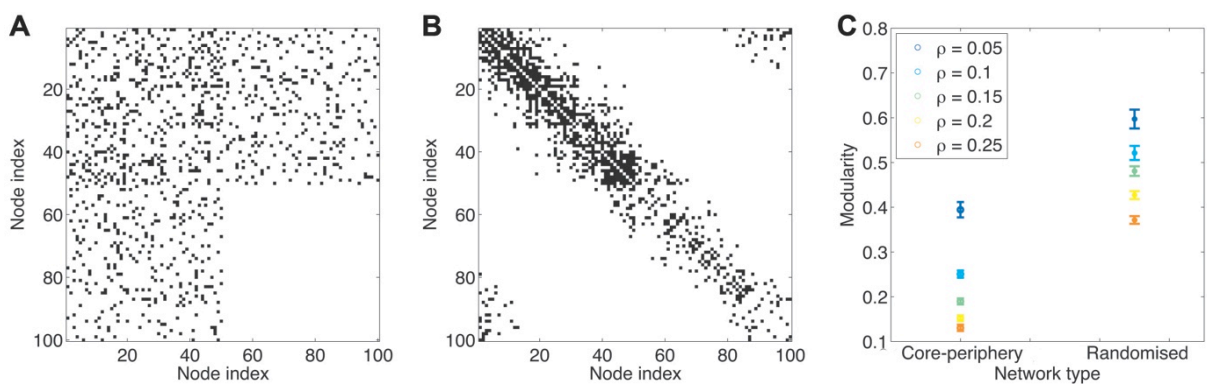

**Figure S6.** Modularity values of synthetic networks. (**A**) Adjacency matrix of a synthetic core-periphery organization consisting of 100 nodes with network density $\rho = 0.1$. (**B**) Adjacency matrix of a network obtained by randomizing the connections of the network presented in panel A while preserving its degree distribution. (**C**) Modularity values of the original and randomized networks for various density levels. Error bars are obtained from 100 realizations for each condition.

In the interaction networks obtained from our experiments, the coevolutionary process is associated with the emergence of core-periphery segregation. Cooperators form strong (i.e., transitive) ties with each other, whereas defectors form weak (i.e., intransitive) ties (*43*) to both cooperators and defectors (Fig. 3C-E, movies S1-4). Due to such asymmetry, modularity values of the interaction networks are significantly lower than those of their random correspondences. As a result, Modularity Z-scores in Fig. 3A exhibit negative values. This result indicates that the modular structure within each interaction network is insignificant when judged by modular structures that are embedded in the corresponding null networks. As illustrated in Fig. 3D, E and movies S1-4, such low values are due to the existence of core-periphery structures. In accordance with this interpretation, maximum modularity values of interaction networks of high cost setting in earlier rounds, in which the segregation process had not yet taken place, and in the later rounds, where the strong end-game effect destroyed the existing core-periphery structure, exhibited relatively small difference with the null values.

On the basis of this finding, we implemented the *k*-shell decomposition method to locate each node in the core-periphery spectrum of the global network structure (see S2 Methods for details). The *k*-shell decomposition method is a classical technique in graph theory that captures structural role differentiation of nodes. The resulting metric, coreness ($k_s$), has been recently rediscovered by applied physicists and has been shown to reveal the structural role differentiation in extremely large empirical networks (*31, 35, 36*) (e.g., the internet) and the dynamic importance of each node in social influence and epidemic spreading (*31*). Unlike the clustering coefficient, which calculates extremely local properties of each node, $k_s$ enables a full specification of nodes in the global topology of a network (*31, 35, 36*). That is, a node with a high clustering coefficient can be located at the periphery if it shares strong links only with other nodes that belong to the periphery (Fig. 3B).

**4.3.2 Toward a Measurement Theory of Social Capital: *K*-Shell Decomposition of Strategic Interaction Networks**

Despite its popularity and theoretical demand in social science research, the notion of social capital has never had a clear-cut definition. The multitude of perspectives that have been proposed by scholars of diverse disciplines has led to an increased ambiguity in the concept of social capital (*44*). Due to its abstract notion, it has been difficult to find a work that measures a direct manifestation of social capital. Rather, existing research attempts to incorporate derivatives that are argued to be consequential products of social capital.

We contend that the *k*-shell decomposition method offers a measurement theory of social capital. We stick to the defining element of the notion that social capital consists of a dense network of cooperative relationships (i.e., a closed network) (*3, 44*). As the series of quantitative analyses in the main text indicates, core structure unveiled by the *k*-shell decomposition method is itself a direct correspondence to this fundamental idea. In contrast to naturally occurring data, which are typically very difficult to utilize for the purpose of elucidating cooperative associations and interaction topology, our stylised laboratory experiments allow us to define such a specific compartment of a strategic interaction network as a closed sub-network.

In addition, the dynamic and multiplex (i.e., multiple types of directed ties: cooperative and non-cooperative) nature of the interaction network data that was obtained from multiple rounds of experiments enables us to trace the generative mechanism of social capital. As depicted in fig. S8, link stability is highly associated with the coreness of each node. Subjects who entered the core structure of the interaction networks were more likely to maintain their previously formed relationships. Given the intimate association between coreness and %C (Fig. 3C), the core's stability also implies that cooperators had stable relationships, in contrast to defectors at the periphery who quickly ended relationships (Fig. 4B, C).

### 4.3.3 Explanatory Power of Degree and Coreness

The $k$-shell structure is a more fundamental topological substrate for the evolution of the system than the number of partners of each node ($k$). The increasing trends of the two main outcomes, %C of players and the cooperators' payoff advantage, are better fit with $k_s$ than $k$. A round index from 1 to 17 exhibits a stronger linear relationship with corr($k_s$, %C) ($r = 0.70\pm0.01$) and corr($k_s$, income) ($r = 0.90\pm0.3$) than with corr($k$, %C) ($r = 0.67\pm0.04$) and corr($k$, income) ($r = 0.79\pm0.17$) (fig. S5A, B).

In addition to $k$-shell decomposition we used a newly proposed measure of coreness (*42*) to compute each subject's coreness level in each round. Despite the fact that these two types of measures intend to characterize different structural indices, they exhibit a fairly high degree of agreement to each other. By Z-scoring each round's coreness vectors obtained using the new measure and $k$-shell decomposition, for 80 networks of the 80 rounds in the 4 sessions, we find that the overall Pearson correlation value of these two types of Z-scored vectors is 0.60 ($p < 10^{-293}$).

## 4.4. Individual Level Strategies

The PD game strategies and networking strategies at the individual level are microscopic mechanisms that underlie the self-organization of social capital. In this section, we review characteristic properties of these strategies.

### 4.4.1. Consistency of the PD Game Strategies at the Individual Level

We often use the words 'cooperators' and 'defectors' to categorise subjects. For this characterisation to be meaningful, subjects' game decisions should exhibit some level of consistency between games in a single round, as well as across rounds of the repeated game. As mentioned in the text, subjects' decisions, measured by %C, showed a high level of temporal stability. Most, if not all, subjects also exhibited consistency in game decisions between games in a round. In any given round, the percentage of subjects who cooperated or defected in all games was 73%, on average. Thus, approximately three quarters of the subjects could be characterized as pure cooperators (All C) or pure defectors (All D) in any given round (table S4). In addition, the subjects who used a consistent PD game strategy increased over time (fig. S7).

**Table S4.** Majority of subjects cooperate or defect in all games of a round. The total numbers of subjects are shown after the session identification numbers. Subject numbers may not add up to the total because in some rounds some subjects had no games to play.

| | LOW COST | | | | | | HIGH COST | | | | | |
|---|---|---|---|---|---|---|---|---|---|---|---|---|
| | **Session 1 (39)** | | | **Session 2 (36)** | | | **Session 1 (35)** | | | **Session 2 (40)** | | |
| Round | All D | Mixed | All C | All D | Mixed | All C | All D | Mixed | All C | All D | Mixed | All C |
| 1 | 12 | 17 | 9 | 16 | 7 | 11 | 7 | 13 | 10 | 10 | 10 | 18 |
| 2 | 16 | 14 | 9 | 14 | 13 | 9 | 11 | 12 | 11 | 13 | 6 | 21 |
| 3 | 16 | 17 | 6 | 12 | 15 | 9 | 9 | 13 | 13 | 9 | 8 | 21 |
| 4 | 17 | 15 | 7 | 14 | 12 | 10 | 9 | 10 | 15 | 10 | 7 | 22 |
| 5 | 14 | 18 | 6 | 13 | 13 | 10 | 5 | 13 | 14 | 9 | 12 | 17 |
| 6 | 13 | 19 | 6 | 13 | 12 | 11 | 7 | 7 | 20 | 8 | 14 | 17 |
| 7 | 12 | 19 | 8 | 8 | 13 | 14 | 9 | 9 | 17 | 9 | 10 | 21 |
| 8 | 13 | 18 | 8 | 12 | 11 | 13 | 6 | 8 | 19 | 9 | 6 | 25 |
| 9 | 14 | 13 | 12 | 10 | 12 | 14 | 10 | 8 | 17 | 9 | 3 | 28 |
| 10 | 12 | 11 | 14 | 8 | 15 | 13 | 7 | 8 | 19 | 8 | 5 | 25 |
| 11 | 13 | 7 | 18 | 10 | 13 | 13 | 6 | 5 | 22 | 8 | 7 | 23 |
| 12 | 12 | 15 | 12 | 8 | 11 | 16 | 5 | 13 | 17 | 9 | 4 | 26 |
| 13 | 10 | 9 | 18 | 8 | 13 | 15 | 6 | 5 | 22 | 10 | 3 | 26 |
| 14 | 10 | 13 | 16 | 9 | 12 | 15 | 8 | 12 | 15 | 7 | 3 | 27 |
| 15 | 11 | 10 | 18 | 9 | 11 | 16 | 7 | 7 | 20 | 6 | 6 | 25 |
| 16 | 11 | 11 | 16 | 8 | 10 | 17 | 5 | 6 | 22 | 10 | 4 | 25 |
| 17 | 11 | 11 | 17 | 9 | 10 | 16 | 6 | 6 | 22 | 10 | 7 | 23 |
| 18 | 12 | 12 | 15 | 8 | 10 | 16 | 8 | 5 | 22 | 10 | 7 | 19 |
| 19 | 10 | 11 | 16 | 13 | 11 | 12 | 11 | 6 | 18 | 19 | 5 | 15 |
| 20 | 23 | 9 | 5 | 30 | 6 | 0 | 24 | 7 | 4 | 31 | 2 | 5 |

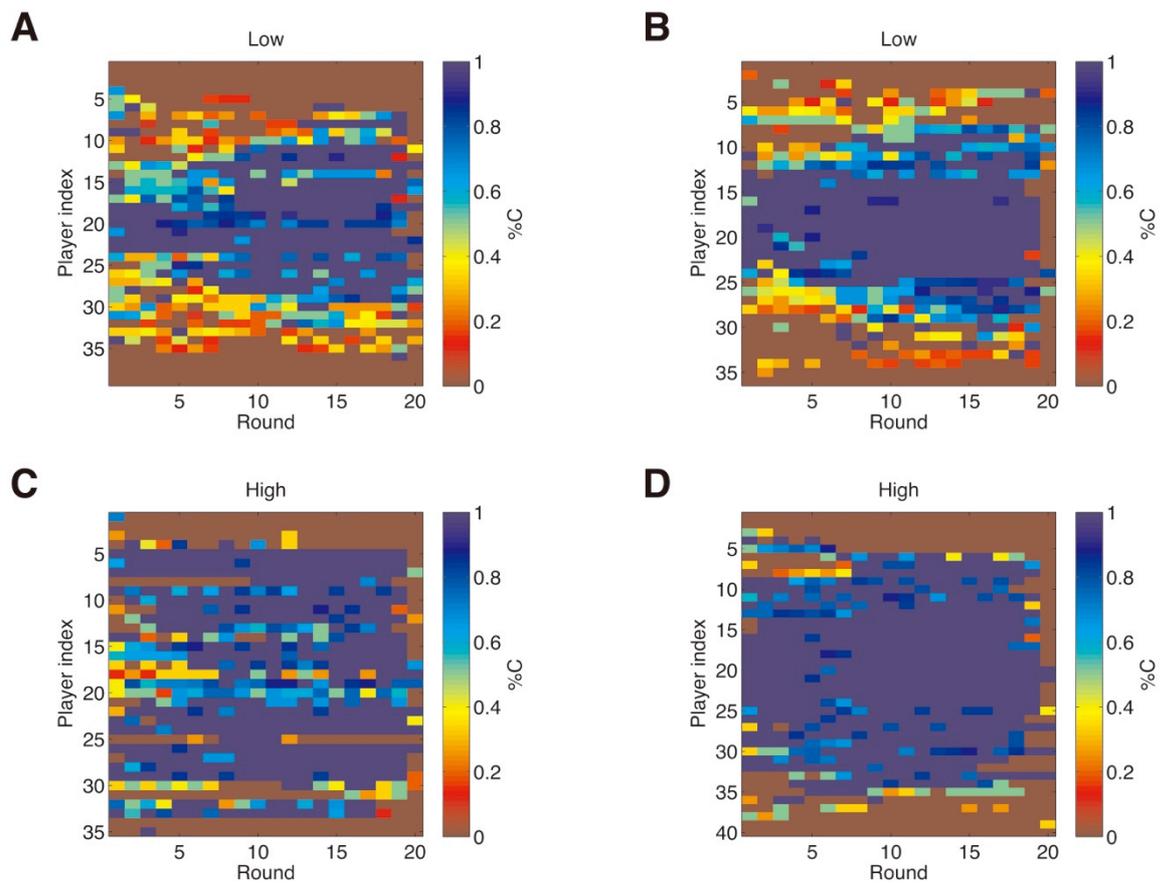

**Figure S7.** Evolution of individual strategic profiles. Temporal stability of individual %C. Each panel represents a session. Color bar indicates %C.

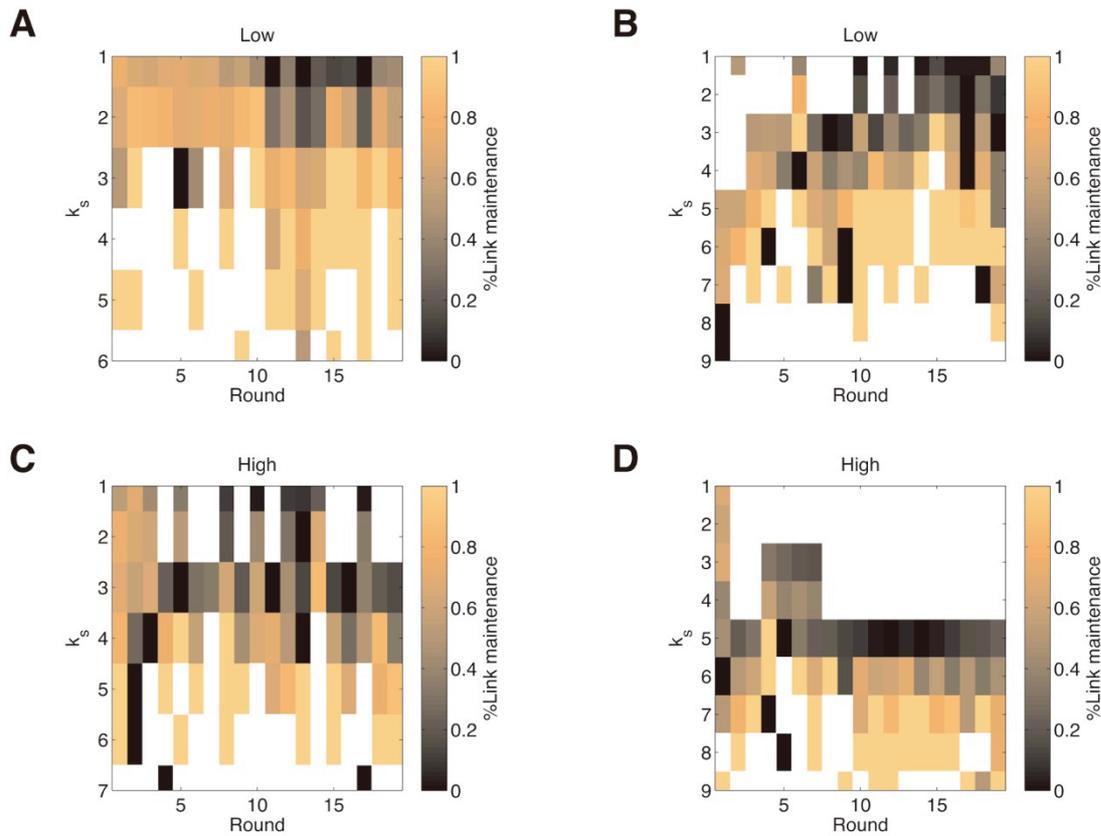

**Figure S8.** Link maintenance probability by subjects' coreness. Percentage of link maintenance in the following round, given each player's measured coreness in the current round.

### 4.4.2 Networking Strategies: Quit-for-tat and Roving

Quit-for-tat and its variants, such as Quit-for-two-tat and forgiving Quit-for-tat, best describe the cooperators' networking strategies, whereas defectors' networking strategies can be characterized as 'Roving.' Table S5 shows networking and PD game strategies of cooperators (those who cooperated in all games in a given round) by setting. The four columns that correspond to each setting are described here. The first column (Partner D) shows the probabilities of proposing in Round t+1 to a Round t partner who defected in the PD game. The second column (Partner C) shows the probabilities of proposing in Round t+1 to a Round t partner who cooperated in the PD game. The third column (Partner D) shows the probabilities of cooperating in Round t+1, conditional on the link continuing to Round t+1, to a Round t partner who defected in the PD game. The fourth column (Partner C) shows the probabilities of cooperating in Round t+1, conditional on the link continuing to Round t+1, to a Round t partner who cooperated in the PD game.

**Table S5.** Cooperators' networking and PD game strategy. The number of cooperators in each round in each setting are shown in table S4.

|       | Low Cost Setting |           |                  |           | High Cost Setting |           |                  |           |
|-------|------------------|-----------|------------------|-----------|-------------------|-----------|------------------|-----------|
|       | Proposal Prob    |           | Cooperation Prob |           | Proposal Prob     |           | Cooperation Prob |           |
| Round | Partner D        | Partner C | Partner D        | Partner C | Partner D         | Partner C | Partner D        | Partner C |
| 1     | 0.27             | 1.00      | 0.14             | 0.95      | 0.08              | 0.95      | 0.00             | 0.93      |
| 2     | 0.27             | 1.00      | 0.00             | 1.00      | 0.33              | 0.97      | 0.50             | 0.98      |
| 3     | 0.00             | 1.00      | .                | 1.00      | 0.34              | 0.99      | 0.50             | 0.93      |
| 4     | 0.14             | 0.99      | 0.00             | 1.00      | 0.37              | 0.96      | 0.80             | 0.99      |
| 5     | 0.11             | 1.00      | 1.00             | 1.00      | 0.74              | 0.99      | 0.71             | 0.99      |
| 6     | 0.15             | 1.00      | 1.00             | 1.00      | 0.47              | 0.99      | 0.64             | 0.99      |
| 7     | 0.43             | 1.00      | 0.78             | 0.99      | 0.29              | 1.00      | 0.67             | 0.99      |
| 8     | 0.55             | 1.00      | 0.89             | 1.00      | 0.50              | 0.99      | 0.67             | 1.00      |
| 9     | 0.54             | 1.00      | 0.56             | 0.99      | 0.48              | 0.99      | 0.80             | 1.00      |
| 10    | 0.56             | 1.00      | 0.40             | 1.00      | 0.58              | 0.99      | 0.50             | 1.00      |
| 11    | 0.24             | 0.99      | 0.50             | 1.00      | 0.33              | 1.00      | 1.00             | 0.99      |
| 12    | 0.52             | 0.99      | 1.00             | 1.00      | 0.44              | 0.99      | 1.00             | 1.00      |
| 13    | 0.50             | 1.00      | 0.00             | 1.00      | 0.62              | 0.99      | 0.83             | 1.00      |
| 14    | 0.53             | 1.00      | 1.00             | 1.00      | 0.19              | 0.99      | 0.50             | 1.00      |
| 15    | 0.79             | 1.00      | 0.00             | 1.00      | 0.20              | 0.99      | 0.00             | 0.98      |
| 16    | 0.35             | 1.00      | 1.00             | 1.00      | 0.80              | 1.00      | 0.25             | 1.00      |
| 17    | 0.33             | 1.00      | 0.67             | 1.00      | 0.50              | 1.00      | .                | 0.93      |
| 18    | 0.40             | 1.00      | 0.50             | 0.78      | 0.32              | 0.99      | 0.00             | 0.86      |
| 19    | 0.51             | 0.96      | 0.00             | 0.43      | 0.61              | 0.99      | 0.00             | 0.38      |

As 'Partner C' columns under Proposal Prob in Table S5 shows, cooperators were all but certain to propose again to others who cooperated and thus, the link almost certainly continued to the next round. In the next round, cooperators almost certainly cooperated again. The cooperators' forgivingness is what appears to be different from the pure Quit-for-tat. Thus, 'Quit-for-two-tat', or 'forgiving Quit-for-tat', most likely better describes the cooperators' networking strategies than pure Quit-for-tat. However, note that the cooperators' probability of proposing to a defecting partner was very low in the first several rounds, and the probability that a cooperator kept the link with a defector for multiple consecutive rounds rapidly approached zero as the game continued. Thus, cooperators quickly accumulated several other cooperators in their ego-centric networks. After that, they could afford to be more forgiving to newly encountered defectors, likely hoping to build additional CC links by converting them to cooperating partners.

**Table S6.** Defectors' networking and PD game strategy. The number of defectors in each round in each setting is shown in table S4.

| | Low Cost Setting | | | | High Cost Setting | | | |
|---|---|---|---|---|---|---|---|---|
| | Proposal Prob | | Cooperation Prob | | Proposal Prob | | Cooperation Prob | |
| Round | Partner D | Partner C | Partner D | Partner C | Partner D | Partner C | Partner D | Partner C |
| 1 | 0.49 | 0.48 | 0.15 | 0.00 | 0.57 | 0.56 | 0.20 | 0.50 |
| 2 | 0.43 | 0.38 | 0.00 | 0.09 | 0.16 | 0.50 | 0.00 | 0.78 |
| 3 | 0.45 | 0.39 | 0.08 | 0.00 | 0.14 | 0.40 | 0.00 | 0.50 |
| 4 | 0.34 | 0.26 | 0.05 | 0.20 | 0.19 | 0.31 | . | 0.67 |
| 5 | 0.26 | 0.32 | 0.07 | 0.33 | 0.13 | 0.38 | 0.00 | 0.75 |
| 6 | 0.29 | 0.54 | 0.22 | 0.14 | 0.32 | 0.61 | 0.00 | 0.08 |
| 7 | 0.07 | 0.34 | 1.00 | 0.25 | 0.23 | 0.54 | 0.00 | 0.00 |
| 8 | 0.23 | 0.57 | 0.30 | 0.00 | 0.39 | 0.54 | 0.00 | 0.00 |
| 9 | 0.23 | 0.41 | 0.00 | 0.17 | 0.20 | 0.39 | 1.00 | 0.00 |
| 10 | 0.21 | 0.52 | 0.00 | 0.17 | 0.26 | 0.35 | 0.00 | . |
| 11 | 0.35 | 0.35 | 0.00 | 0.25 | 0.45 | 0.57 | 0.00 | 1.00 |
| 12 | 0.32 | 0.29 | 0.00 | 0.00 | 0.18 | 0.50 | 0.00 | 0.00 |
| 13 | 0.18 | 0.50 | 0.00 | 0.00 | 0.23 | 0.40 | 0.00 | 0.00 |
| 14 | 0.17 | 0.44 | 0.25 | 0.00 | 0.19 | 0.50 | 0.00 | 0.00 |
| 15 | 0.27 | 0.25 | 0.00 | 0.00 | 0.19 | 0.86 | . | 0.00 |
| 16 | 0.16 | 0.36 | 0.00 | 0.00 | 0.30 | 0.44 | 0.00 | 0.00 |
| 17 | 0.15 | 0.31 | 0.00 | 0.00 | 0.41 | 0.20 | 0.00 | . |
| 18 | 0.12 | 0.26 | 0.00 | 0.00 | 0.46 | 0.31 | 0.00 | 0.00 |
| 19 | 0.31 | 0.59 | 0.00 | 0.00 | 0.35 | 0.71 | 0.00 | 0.00 |

Table S6 shows similar statistics for the defectors in each round. Several aspects of the data confirm that 'Roving' was the most common networking strategy among the defectors, and this was increasingly the case as the game continued. First, defectors very often did not propose to a cooperator whom the defector exploited in the previous round. The probability that a defector would propose again to a cooperator in the next round seldom rose above 0.5 in the low cost setting, and the probability fell below 0.5 in the majority of rounds in the high cost setting. These low probabilities are a bit puzzling. We conjecture that the defectors themselves knew that cooperators were not likely to be pure altruists, but were conditional cooperators. They most likely also knew that cooperators were not likely to propose to them in the next round, and given the limited number of proposals one can make in a round, proposing to a once-exploited cooperator could waste one's proposals. Even if the link continued by mutual proposals, the defectors most likely thought that the likelihood of the partner cooperating again was very low, in which case the expected outcome was DD.

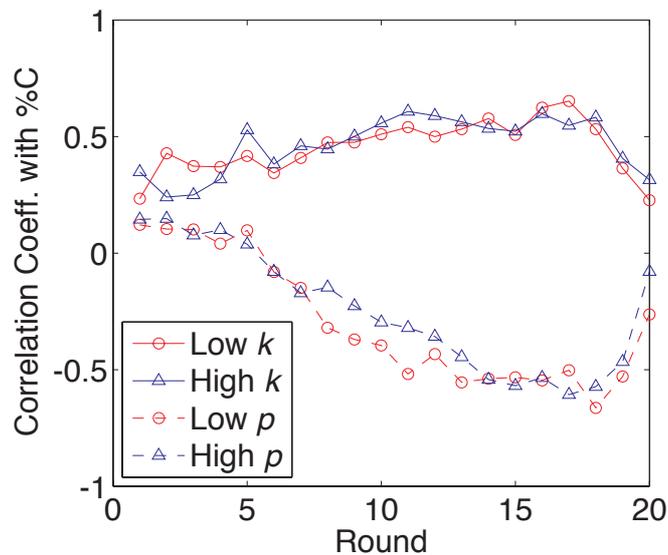

**Figure S9.** Correlations between the number of proposals (and accepted proposals) and %C. Effect of %C on the number of accepted proposals ($k$) and the number of proposals ($p$). Solid lines indicate correlation coefficients between %C and $k$. Dashed lines represent correlation coefficients between %C and $p$.

The high probabilities of cutting partners of both types and the near zero probability of cooperating with a partner when matched again show that the defectors attempted to find new

"suckers" every round. However, because most cooperators were Quit-for-taters and, more importantly, stopped searching for new partners once enough CC links were established, the roving defectors were doomed to either get rejected by cooperators or to get matched with other defectors. The narrowing scope of the cooperators' search for new partners and the desperate, but unsuccessful, attempts by defectors in search of new suckers, are shown by the correlation coefficients between individual's cooperation rate (%C) and the number of proposals ($p$) and the number of accepted proposals ($k$) (fig. S9). Defectors made more proposals but had less links than cooperators. This trend grew stronger over time.

## 5 Supplementary Movies

**Movie S1 (mp4)**
Session 1 of the low cost setting

**Movie S2 (mp4)**
Session 2 of the low cost setting

**Movie S3 (mp4)**
Session 1 of the high cost setting

**Movie S4 (mp4)**
Session 2 of the high cost setting

Supplementary Movies depict the coevolution of individual game strategy and network structure by connecting network images of multiple rounds of each session. For smoothing the connected layers of different rounds, we utilized multi-layer Kamada-Kawai algorithm in a social network animation program, SoNIA (http://sonia.stanford.edu). Numbers in the upper left, after "time," correspond to the index of the game round depicted in the frame (e.g., 1.3-2.3 means that the image is a result of smoothing interaction structures of rounds 1 and 2). The color of each node indicates %C of the depicted player in the round. The rate of cooperation in a given round increases as a node becomes blue and decreases as a node becomes red. See the color bar in Fig. 3e for the reference color scale. A blue arc with an arrow indicates the originator's cooperative decision to its receiver, whereas an arc in red represents defection. Nodes are distributed in a spring-embedded layout, which locates tightly connected nodes in proximal distances. A bigger node indicates higher $k_s$. Densely connected nodes with high $k_s$ values are located at the centre of the network, whereas sparsely connected nodes are located at the periphery, separated from the rest of the network. These animations capture the emergence of positive assortment of similar types, followed by core-periphery segregation, from an initial well-mixed structure of cooperators and defectors (Round 1). Around round 15~17, cooperators form a dense and closed network at the centre, whereas defectors form loosely connected, open networks at the periphery in later rounds. The collapse of core-periphery structure and the decrease in rate of cooperation in the final round capture strong end game effects.

# References

32. V. D. Blondel, J. L. Guillaume, R. Lambiotte, E. Lefebvre, Fast unfolding of communities in large networks. *Journal of Statistical Mechanics: Theory and Experiment* **2008**, P10008 (2008).
33. R. Guimera, M. Sales-Pardo, L. A. N. Amaral, Modularity from fluctuations in random graphs and complex networks. *Physical Review E* **70**, 025101 (2004).
34. S. Maslov, K. Sneppen, Specificity and stability in topology of protein networks. *Science* **296**, 910 (2002).
35. S. Carmi, S. Havlin, S. Kirkpatrick, Y. Shavitt, E. Shir, A model of internet topology using k-shell decomposition. *Proceedings of the National Academy of Sciences* **104**, 11150 (2007).
36. M. Kitsak, M. Riccaboni, S. Havlin, F. Pammolli, H. E. Stanley, Scale-free models for the structure of business firm networks. *Physical Review E* **81**, 036117 (2010).
37. N. Hayashi, From Tit-for-Tat to Out-for-Tat. *Sociological Theory and Methods* **8**, 19 (1993).
38. N. Kashtan, U. Alon, Spontaneous evolution of modularity and network motifs. *Proceedings of the National Academy of Sciences* **102**, 13773 (2005).
39. D. S. Bassett *et al.*, Core-Periphery Organisation of Human Brain Dynamics. *arXiv preprint arXiv:1210.3555*, (2012).
40. R. S. Burt, J. Merluzzi, in *Research in the Sociology of Organizations,* S. P. Borgatti, D. J. Brass, D. S. Halgin, G. Labianca, A. Mehra, Eds. (Emerald Group Publishing, Cambridge, MA, 2013).
41. P. S. Dodds, D. J. Watts, C. F. Sabel, Information exchange and the robustness of organizational networks. *Proceedings of the National Academy of Sciences* **100**, 12516 (2003).
42. M. P. Rombach, M. A. Porter, J. H. Fowler, P. J. Mucha, Core-Periphery Structure in Networks. *arXiv preprint arXiv:1202.2684*, (2012).
43. M. S. Granovetter, The strength of weak ties. *American journal of sociology*, 1360 (1973).
44. N. Lin, K. S. Cook, R. S. Burt, *Social capital: theory and research*. (Aldine de Gruyter, 2001).